\documentclass[a4paper,10pt]{article}
\usepackage[utf8]{inputenc}

\usepackage{amsmath,amssymb,amsthm,latexsym}

\newcommand{\bea}{\begin{eqnarray}}
\newcommand{\eea}{\end{eqnarray}}
\newcommand{\bee}{\begin{equation}}
\newcommand{\ee}{\end{equation}}

\newcommand{\Tr}{{\rm Tr}}

\title{Large-N dynamics of the spiked tensor model with random
initial conditions}

\begin{document}

\author{Vasily Sazonov\footnote{vasily.sazonov@cea.fr}\\
Universit\'e  Paris-Saclay,  CEA,  List,\\  \textit{F-91120,  Palaiseau,  France}}

\maketitle

\begin{abstract}
In these notes, we develop a path integral approach for the partial differential equations with random initial conditions. Then, we apply it to the dynamics of the spiked tensor model and show that the large-$N$  saddle point equations are dominated by the melonic type diagrams.
\end{abstract}

\section{Introduction}
Non-convex multidimensional optimization and the related problem of finding the global minimum in rough landscapes are crucial challenges of modern science. 
Such problems were extensively studied in the context of the spin-glass systems \cite{SGrev}, and found applications in biology \cite{BioBook}, finance \cite{finance}, and data science \cite{InfoBook}. Here, we focus on a task motivated by data science and consider the model of the signal recovering from a noisy high-dimensional data tensor -- the spiked tensor model (tensor PCA) \cite{tensorPCA0, tensorPCA, Ros}. 
In tensor PCA, the data ${Y} \in ({\mathbb{R}}^N)^{\otimes p}$ is assumed to be of the form
\begin{equation}
  {Y} = \lambda {u}^{\otimes p} + {J}\,,
\label{SpEq}
\end{equation}
where $\lambda$ is a relative strength of the signal, $u$ is a spherically normalized, $\sum_{i} u^2_i = N$, vector (a subject for the estimation), and $J \in ({\mathbb{R}}^N)^{\otimes p}$ is a normally distributed noise tensor and one is generally interested in the large-$N$ limit.

According to the analysis, \cite{tensorPCA0, tensorPCA} the spiked tensor model exhibits three regimes: for $\lambda < C \sqrt{p \log p}$ there is no statistical difference between the presence and absence of the signal, for $\lambda > C N^{(p-1)/4}$ the signal can be recovered in polynomial time, and no efficient algorithms are known in the intermediate mode.

Optimization algorithms used in machine learning and statistical inference are frequently based on (stochastic) gradient descent. In the recent work, \cite{LangevinSpTM} the Langevin algorithm applied to the tensor PCA problem was investigated. It was shown there that the algorithmic signal-to-noise threshold of the standard Langevin algorithm is sub-optimal compared to the one given by approximate message passing (AMP).
However, the success of the gradient-descent-based optimization may depend on the initial approximation, therefore the utilization of the different starting points (or averaging with respect to them) can in principle improve the performance of the Langevin algorithm. Thus, the study of the Langevin algorithm with the random starting points (initial conditions) is of great interest. Here we do a modest step in this direction by focusing on the gradient dynamics without the Langevin force term but with the random initial conditions and derive a closed set of equations for these dynamics in the large-N limit. We show that these equations exhibit the dominance of the melonic diagrams, typical for the random tensor models \cite{Gurau}.
In the following section, we derive a supersymmetric path integral representation for the gradient-type PDEs with random initial conditions. Then, in section \ref{sptm} we apply the developed approach to the spiked tensor model.

\section{Path integral representation for the gradient-type PDEs with random initial conditions}
\label{QFT}
Consider an equation
\begin{eqnarray}
  \partial_t \phi_i(t) = -\frac{\delta H}{\delta \phi_i}(t)\,,~~~\phi_i(t_0) = \varphi_i\,,
\label{Eq1}
\end{eqnarray}
where $\frac{\delta H}{\delta \phi_i}(t)$ consists of linear and non-linear terms, and the statistic of initial conditions is given by
\begin{eqnarray}
\label{Cij}
  C_{ij} = \langle \varphi_i\varphi_j\rangle_\varphi = \frac{\delta_{ij}}{N}\chi_{N}(j), \\
  \sum_{j = 1}^N \chi_N(j) = N\,.
\label{norm}
\end{eqnarray}
According to the chosen covariance, the components of the initial conditions vector are distributed independently but not necessarily identically. In models, where the field can take arbitrary values, condition \eqref{norm} may be relaxed.
In light of the following application of this approach to the spiked vector model, it reflects a possibility that some prior information about the recovering signal is known.
The conditions \eqref{Cij} and \eqref{norm} lead to an obvious spherical constraint for the averages of the initial conditions,
\begin{eqnarray}
  \sum_{j=1}^N \langle\varphi_j^2\rangle = 1\,.
\end{eqnarray}

To derive a path integral representation for \eqref{Eq1}, we, first of all, equivalently rewrite \eqref{Eq1} in the integral form as
\begin{eqnarray}
  \phi_i(t) =\varphi_i - \int_{t_0}^t\,d\tau\,\frac{\delta H}{\delta \phi_i}(\tau)\,.
\label{Eq2}
\end{eqnarray}
The latter form has the advantage that the initial conditions are now explicitly present in the equation. Starting from this point we can apply the procedure similar to the Faddeev-Popov gauge fixing \cite{FP} or derivation of the Martin-Siggia-Rose effective action \cite{MSR}.

Assume that $\vec{\phi^0}(t)$ is a solution of the equation \eqref{Eq2}, then we can formally write an average of some operator $O[\vec{\phi^0}]$
with respect to the random initial conditions as
\begin{eqnarray}
  \big\langle O(\vec{\phi^0})\big\rangle_\varphi = \Big\langle \int D\vec\phi\, O[\vec\phi]\, \delta(\vec\phi - \vec{\phi^0})\Big\rangle_\varphi\,,
\label{phi_av1}
\end{eqnarray}
where the delta function of a vector argument stands for the product of delta functions of vector components, $\delta(\vec\phi - \vec{\phi^0}) = \prod_{i=1}^N \delta(\phi_i - \phi_i^0)$, and
\begin{eqnarray}
  &&\langle ... \rangle_\varphi = \frac{\int D\vec\varphi\,... e^{-\varphi_i C_{ij}^{-1} \varphi_j/2}}{\int D\vec\varphi\, e^{-\varphi_i C_{ij}^{-1} \varphi_j/2}}\,.
\label{av}
\end{eqnarray}
Then, we rewrite the delta function as,
\begin{eqnarray}
    &&\delta(\vec\phi - \vec{\phi^0}) = \det M \cdot \prod_{i=1}^N \delta(\phi_i(t) - \varphi_i + \int_{t_0}^t\,d\tau\,\frac{\delta H}{\delta \phi_i}(\tau))\,,\\
    &&M_{ij}(t, t') \equiv \frac{\delta}{\delta\phi_j(t')}\Big(\phi_i(t) - \varphi_i + \int_{t_0}^t\,d\tau\,\frac{\delta H}{\delta \phi_i}(\tau)\Big)\,.
\end{eqnarray}
Employing the Fourier representation of the delta function and using the fermionic ghost fields to write a determinant term, we end up with
\begin{eqnarray}
  \big\langle O(\vec{\phi^0})\big\rangle_\varphi=\big\langle\int D\vec\phi\, \int D\vec{\hat\phi}\, \int D\vec{\bar\xi}\,\int D\vec\xi\, 
  O[\vec\phi]\, e^{-S_1}\big\rangle_\varphi\,,
\end{eqnarray}
where
\begin{eqnarray}
  S_1 &:=& \sum_{i=1}^N\int_{t_0}^{t_e} dt\, \Big(-i\hat\phi_i(t)\big[-\phi_i(t) - \int_{t_0}^t d\tau\, \frac{\delta H}{\delta \phi_i}(\tau) + \varphi_i\big]\Big)\nonumber\\
  &-& \sum_{i,j=1}^N\int_{t_0}^{t_e} dt\,\int_{t_0}^{t_e} dt'\,\bar\xi_i(t)\Big[\delta_{ij}\delta(t - t')\nonumber\\
  &+& \int_{t_0}^t d\tau\,\frac{\delta}{\delta \phi_j(t')}\frac{\delta H}{\delta \phi_i}(\tau)\Big]\xi_{j}(t')\,.
\label{S1}
\end{eqnarray}
Performing the variation $\frac{\delta}{\delta \phi_j(t')}$ in the term with ghost fields $\bar\xi_i$, $\xi_i$, we simplify it as
\begin{eqnarray}
  &&\int_{t_0}^{t_e} dt\,\int_{t_0}^{t_e} dt'\,\bar\xi_i(t)\Big[\delta_{ij}\delta(t - t') + \int_{t_0}^t d\tau\,\delta(\tau - t') \frac{\delta^2 H}{\delta \phi_i \delta \phi_j}(\tau)\Big]\xi_{j}(t')\nonumber\\
  &&= \int_{t_0}^{t_e} dt\,\int_{t_0}^{t} dt'\,\bar\xi_i(t)\Big[\delta_{ij}\delta(t - t') + \frac{\delta^2 H}{\delta \phi_i \delta \phi_j}(t')\Big]\xi_{j}(t')\,.
\label{S1_2}
\end{eqnarray}
Let us note, that contrary to the Langevin dynamics, the considered equations with random initial conditions describe systems with memory, due to the terms containing $\int_{t_0}^t dt...$ .

Averaging over initial conditions, we obtain
\begin{eqnarray}
  \langle O(\vec{\phi^0})\rangle_\varphi = 
  \int D\vec\phi\, \int D\vec{\hat\phi}\, \int D\vec{\bar\xi}\,\int D\vec\xi\, 
  O[\vec\phi]\, e^{-S_2}\,
\label{O2}
\end{eqnarray}
with
\begin{eqnarray}
  S_2 &:=& \sum_{i = 1}^N \Bigg[\frac{\chi_N(i)}{2N}\Big(\int_{t_0}^{t_e} dt\, \hat\phi_i(t) \Big)^2 + 
  i\int_{t_0}^{t_e} dt\, \hat\phi_i(t)\phi_i(t)\nonumber\\
  &+& i\int_{t_0}^{t_e} dt\,\int_{t_0}^t dt'\, \hat\phi_i(t) \frac{\delta H}{\delta \phi_i}(t')\Bigg]\nonumber\\
  &-& \sum_{i,j = 1}^N\int_{t_0}^{t_e} dt\,\int_{t_0}^{t} dt'\,\bar\xi_i(t)\Big[\delta_{ij}\delta(t - t') + \frac{\delta^2 H}{\delta \phi_i \delta \phi_j}(t')\Big]\xi_{j}(t')\,.
\label{S2}
\end{eqnarray}
Let us define a superfield, bi-local in time -- the standard choice of the superfield in the Langevin dynamics contains only a one-time variable,
\begin{equation}
  \Phi_i(\bar\theta, \theta, t, t') := \phi_i(t') + \bar\theta\xi_i(t') + \bar\xi_i(t) \theta + i\hat\phi_i(t)\bar\theta\theta\,,
\label{defPhi}
\end{equation}
the corresponding integration in the superspace is defined as
\begin{equation}
  \int d1 := \int_{t_0}^{t_e} dt_1\,\int_{t_0}^{t_1} dt_1'\,\int d\theta_1\,\int d\bar\theta_1\,.
\end{equation}
Then, the first term of \eqref{S2} is given by
\begin{equation}
  \sum_{i = 1}^N \frac{\chi_N(i)}{2N} \int d1\,\int d2\, \frac{1}{(t_1-t_0)(t_2-t_0)} \Phi_i(1)\Phi_i(2)\,,
\end{equation}
and for the sum of the second term and the term with $\bar\xi_i(t)\delta_{ij}\delta(t - t')\xi_i(t')$ we have
\begin{eqnarray}
  &&\int d1\,\Phi_i(1) \delta(t_1 - t_1') \theta_1 \partial_{\theta_1} \Phi_i(1)\nonumber\\
  &&= \int d1\, \delta(t_1 - t_1') \big[\phi_i(t_1')\theta_1 + \bar\theta_1 \xi_i(t_1')\theta_1\big]\big[-\bar\xi_i(t_1) - i\hat\phi_i(t_1')\bar\theta_1\big]\nonumber\\
  &&= \int d1\, \delta(t_1 - t_1') \big[i\hat\phi_i(t_1')\phi_i(t_1')\bar\theta_1\theta_1 - \bar\xi_i(t_1)\xi_i(t_1') \bar\theta_1\theta_1\big]\,.
\end{eqnarray}
The sum of two other terms in \eqref{S2} is equal to
\begin{equation}
  \int d1\, H(\Phi(1))\,.
\end{equation}

The model defined by the action \eqref{S2} obeys supersymmetry of the BRST type. Namely, \eqref{S2} is invariant under an infinitesimal shift in the variable $\bar\theta \rightarrow \bar\theta + \epsilon$, which results in the following change in the superfield $\Phi$:
\begin{eqnarray}
    \Phi \rightarrow \Phi_i + \epsilon\, \partial_{\bar\theta}\Phi_i
    = \Phi_i + \epsilon\,(\xi_i(t') + i\hat\phi_i(t)\theta)\,.
\label{transf}
\end{eqnarray}
The transformation \eqref{transf} is bi-local in time, however, this doesn't prohibit supersymmetry, and the corresponding invariance of the action and measure can be easily verified by direct substitution.

\section{Spiked tensor model}
\label{sptm}
Here we consider the spiked tensor model and employing the path integral representation from the previous section, derive the large-$N$ saddle point equations for its dynamics with random initial conditions.
The Hamiltonian of the spiked tensor mode is defined as
\begin{eqnarray}
  H &:=& -\sum_{i_1<..<i_p}J_{i_1..i_p} \cdot \phi_{i_1}\phi_{i_2}\cdot...\cdot \phi_{i_p}\nonumber\\ 
  &-& \frac{r}{2 N^{p-1}}\sum_{i_1,..,i_p} x_{i_1}\cdot...\cdot x_{i_p}\cdot \phi_{i_1}\cdot...\cdot \phi_{i_p}
  + \mu \big(\sum_i^N\phi_i^2 - N\big)\,,
\label{H}
\end{eqnarray}
where $x_{i}$ is a signal and $\frac{r}{2 N^{p-1}} \equiv \lambda$ is its strength, 
the signal obeys $\sum_{i=1}^N x^2_i = N$, and deviations from the spherical constrain for the field $\sum_{i=1}^N \phi^2_i = N$ are penalized by the term proportional to some large $\mu$.
The tensor $J_{i_1..i_p}$ represents a symmetric noise with Gaussian statistics
\begin{equation}
\overline{J^2_{i_1..i_p}} = \frac{J^2 p!}{2 N^{p-1}}\,.
\label{Jcov}
\end{equation}
%
Substituting the Hamiltonian \eqref{H} to the formulas from the previous section, for an average of an operator $O$, we obtain
\begin{eqnarray}
  \langle O \rangle &=&\int D \Phi\,O\, \exp\Big[-\int d1\,\int d2\, \sum_{i,j = 1}^N \Phi_i(1) K_{ij}(1, 2)\Phi_j(2)\nonumber\\
  &+& H_J[\Phi] + H_S[\Phi] - H_\mu[\Phi]\Big]\,,
\label{avO}
\end{eqnarray}
where 
\begin{eqnarray}
  K_{ij}(1, 2) &:=&
  \delta_{ij}\Big[\frac{1}{(t_1-t_0)(t_2-t_0)} \frac{\chi_N(i)}{2N} \nonumber\\
  &+&\, \delta(1 - 2) \delta(t_1 - t_1') \theta_1\partial_{\theta_2}\Big]\,,
\end{eqnarray}
$H_J$ is the disordered Hamiltonian
\begin{equation}
  H_J := \int d1 \sum_{i_1<..<i_p} J_{i_1..i_p} \Phi_{i_1}(1)\Phi_{i_2}(1)\cdot...\cdot \Phi_{i_p}(1)\,,
\end{equation}
the term $H_S$ contains the signal part,
\begin{eqnarray}
  H_S &:=& \frac{r}{2 N^{p-1}} \int d1 \sum_{i_1,..,i_p}  x_{i_1}x_{i_2}\cdot...\cdot x_{i_p} \Phi_{i_1}\Phi_{i_2}(1)\cdot...\cdot \Phi_{i_p}(1) \nonumber\\
  &=& \frac{r}{2} N \int d1 \Big(\frac{1}{N}\sum_{i}x_{i} \Phi_{i}(1)\Big)^p\,,
\end{eqnarray}
and
\begin{eqnarray}
  H_\mu &:=& \mu \int d1 \big(\sum_{i} \Phi_{i}^2(1) - N\big)\,.
\end{eqnarray}
Integrating out the disorder leads to the substitution of the term $H_J$ by 
\begin{eqnarray}
  H_{\int J} := \frac{J^2 p!}{2 N^{p-1}} \int d1 \int d2 \sum_{i_1<..<i_p} \Phi_{i_1}(1)\Phi_{i_1}(2)\cdot...\cdot \Phi_{i_p}(1)\Phi_{i_p}(2)\,.
\end{eqnarray}
In the expression above the restricted sum over $i_1<..<i_p$ in the large-$N$ limit can be replaced \cite{KurchanSUSY, Sommers} by
\begin{equation}
  \frac{p!}{2 N^{p-1}} \sum_{i_1<..<i_p}... = \frac{1}{2 N^{p-1}} \Big(\sum_{i_1,..,i_p}... - \frac{p(p-1)}{2}\sum_{i_1,i_1\neq i_3..,i_p}...\Big) + O(1/N^2)\,.
\label{simpl}
\end{equation}
Now we introduce two collective fields -- pair correlation and response functions,
\begin{eqnarray}
  Q(1, 2) = \frac{1}{N}\sum_{i} \Phi_{i}(1)\Phi_{i}(2)\,,\\
  R(1, 2) = \frac{1}{N}\sum_{i} x_i \Phi_{i}(1)\,,
\end{eqnarray}
using the following identities
\begin{eqnarray}
  1 &=& \int DQ\int D\hat{Q} \exp\Big[\frac{1}{2}\int d1\int d2\, \big(N Q(1, 2) \hat{Q}(1, 2)\nonumber\\ 
  &-& \hat{Q}(1, 2) \sum_i \Phi_i(1)\Phi_i(2)\big)\Big]\,,\\ 
  1 &=& \int DR\int D\hat{R} \exp\Big[\int d1\, \big(N R(1) \hat{R}(1) - \hat{R}(1) \sum_i x_i \Phi_i(1)\big)\Big]\,. 
\end{eqnarray}
The second sum in \eqref{simpl} is suppressed by the factor $1/N$, thus the expectation value \eqref{avO} can be written as
\begin{eqnarray}
  \langle O \rangle &=& \int D\Phi\,DQ\,D\hat{Q}\,DR\,D\hat{R}\,O\, \exp\Big[-\int d1\,\int d2\, \nonumber\\
  &&\Big\{\sum_{i,j = 1}^N \Phi_i(1) \big[K_{ij}(1, 2) + \frac{\delta_{ij}}{2}\hat{Q}(1, 2)\big]\Phi_j(2)\nonumber\\
  &-& \frac{N}{2} \big[J^2 Q^{\text{\textbullet} p}(1, 2) + \hat{Q}(1, 2) Q(1, 2)\big]\Big\}\nonumber\\
  &+& N \int d1\, \big[\frac{r}{2} R^p(1) + \hat{R}(1) R(1)\big] - \int d1\, \hat{R}(1) \sum_i x_i \Phi_i(1) \nonumber\\
  &-& \mu N \int d1\,\big[Q(1, 1) - 1\big]\Big]\,,
\label{avO2}
\end{eqnarray}
where the sign \textbullet\, indicates ordinary multiplication, non-convolution.
Integrating out initial degrees of freedom, making use of $\sum_i^N x_i^2 = N$, and performing a shift
\begin{equation}
  \bar{Q}(1, 2)/2 := \frac{1}{N}\Big[\sum_{i,j}K_{ij}(1, 2) + \frac{\delta_{ij}}{2}\hat{Q}(1, 2)\Big] =: \frac{1}{2}\Big[K(1, 2) + \hat{Q}(1, 2)\Big]\,,
\end{equation}
we obtain
\begin{eqnarray}
  \langle O \rangle &\sim& \int DQ\,D\bar{Q}\,DR\,D\hat{R}\, \widetilde{O}\, \exp\Big[-\frac{N}{2}\int d1\,\int d2\,\nonumber\\
   &&\big[(K(1, 2)-\bar{Q}(1, 2)) Q(1, 2)\nonumber\\
  &-& J^2 Q^{\text{\textbullet} p}(1, 2)\big]
  + N \int d1\, \big[\frac{r}{2} R^p(1) - \hat{R}(1) R(1)\big]\nonumber\\
  &+& \frac{N}{2} \int d1\,d2\, \hat{R}(1)\bar{Q}^{-1}(1, 2)\hat{R}(2)
  +\frac{N}{2}\Tr\log\bar{Q}(1, 2)\nonumber\\
  &-& \mu N \int d1\,\big[Q(1, 1) - 1\big]\Big]\,.
\label{avO3}
\end{eqnarray}
Then, integrating over $\hat{R}$, we get
\begin{eqnarray}
  \langle O \rangle &\sim& \int DQ\,D\bar{Q}\,DR\, \widetilde{O}\, \exp\Big[-\frac{N}{2}\int d1\,\int d2\,\nonumber\\
   &&	\big[(K(1, 2)-\bar{Q}(1, 2)) Q(1, 2) - J^2 Q^{\text{\textbullet} p}(1, 2)\big]\nonumber\\
  &+& \frac{N}{2} r \int d1\, R^p(1)
  - \frac{N}{2} \int d1\,d2\, R(1)\bar{Q}(1, 2) R(2) \nonumber\\
  &+&(\frac{N}{2} - \frac{1}{2})\Tr\log\bar{Q}(1,2)
  - \mu N \int d1\,\big[Q(1, 1) - 1\big]\Big]\,.
\label{avO4}
\end{eqnarray}
The saddle-point equations can be obtained by performing variations of the action with respect to $\bar Q(1,2)$, $Q(1, 2)$, and $R(1)$ as
\begin{eqnarray}
  &&Q(1, 2) - R(1)R(2) + \bar{Q}^{-1}(1, 2) = 0\,,\\
  &&-K(1, 2) + \bar{Q}(1, 2) + p\,J^2 Q^{\text{\textbullet} (p-1)}(1, 2) = 0\,,\\
  &&r\, p\, R^{p-1}(1) - \int d2\, \bar{Q}(1, 2) R(2) = 0\,.
\end{eqnarray}
The large parameter $\mu$ fixes the spherical constraint as
\begin{equation}
    Q(1, 1) = 1\,.
\end{equation}
Substituting $\bar{Q}(1, 2) = (R(1)R(2) - Q(1, 2))^{-1}$ in the last two equations, we end up with
\begin{eqnarray}
  &&-K(1, 2) + (R(1)R(2) - Q(1, 2))^{-1}
  + p\,J^2 Q^{\text{\textbullet} (p-1)}(1, 2) = 0\,,\\
  &&r\, p\, R^{p-1}(1) + \int d2\, (Q(1, 2) - R(1)R(2))^{-1} R(2) = 0\,.
\end{eqnarray}

\section{Concluding remarks}
For all signal strengths $r$, there exists a solution of equations above such that $R(1) = 0$ and $Q(1, 2)$ obeys the saddle point equation of the $p$-spin model. This equation is dominated by the diagrams of the melonic type standard for the large-$N$ limit of the spin glasses and random tensor models \cite{Gurau, Cugliandolo}.
When $\lVert R(1)R(2)\rVert \gg \lVert Q(1, 2)\rVert $, the initial approximation for the $R(1)$ function is determined again by the melonic type equation
\begin{eqnarray}
  r\, p\, R^{p-1}(1) - \int d2\, (R(1)R(2))^{-1} R(2) = 0\,,
\end{eqnarray}
equivalent to 
\begin{equation}
  \int d1\, R^p(1) = \frac{1}{r p}\,.
\end{equation}
The equations above indicate two general possibilities for the solutions: zero and non-zero overlap with the recovering vector.
To derive ultimate saddle point equations, we have extended the super-field formalism of stochastic differential equations to the case of differential equations with random initial conditions, this approach  might be useful for many other applications, including, for instance, the analysis of the melonic turbulence \cite{Turb}.

\end{document}